\documentclass[preprint,12pt]{elsarticle}

\usepackage{amssymb}

\journal{Astroparticle Physics}

\begin{document}

\begin{frontmatter}



\title{Sommerfeld enhancement of invisible dark matter annihilation in
galaxies and galaxy clusters}


\author{Man Ho Chan}

\address{Department of Science and Environmental Studies, The Hong Kong 
Institute of Education}

\begin{abstract}
Recent observations indicate that core-like dark matter structures exist
in many galaxies, while numerical simulations reveal a singular
dark matter density profile at the center. In this article, I show that if
the annihilation of dark matter particles
gives invisible sterile neutrinos, the Sommerfeld enhancement of the
annihilation cross-section can give a sufficiently large annihilation rate
to solve the
core-cusp problem. The resultant core density, core radius, and their
scaling relation generally agree with recent empirical fits from
observations. Also, this model predicts
that the resultant core-like structures in dwarf galaxies can be easily
observed, but not for large normal galaxies and galaxy clusters.
\end{abstract}


\end{frontmatter}


\section{Introduction}
It is commonly believed that the existence of dark matter can account for
the missing mass in galaxies, galaxy clusters and our universe. However,
the nature of dark matter remains a fundamental problem in astrophysics.
If the dark matter particles are cold and collisionless, $N$-body
simulations show that the density profile should be singular at the center
(a cusp profile, $\rho \sim r^{-1}$) \cite{Navarro}. This model generally gives good agreements with observations on
large-scale structures such as Ly$\alpha$ spectrum \cite{Croft,Spergel}
and some galaxy clusters \cite{Pointecouteau}. However,
observations reveal that cores exist in many galaxies ($\rho \sim r^{-\gamma}$ with $\gamma <0.5$), especially in dwarf
galaxies \cite{Salucci,Borriello,Zackrisson}. Some dwarf galaxies can even have $\gamma < 0.2$ \cite{deBlok}. This
discrepancy is commonly known as the core-cusp problem \cite{deBlok}.
Moreover, recent studies show that this problem might also be assoicated 
with another problem, called the too-big-to-fail (TBTF) problem. This   
problem illustrates the fact that the densities of dark matter subhaloes
which surround nearby dwarf spheroidal galaxies are significantly lower
than those of the most massive subhaloes expected around a normal sized   
galaxies in cosmological simulations \cite{Boylan,Boylan2}. In other
words, solving the core-cusp problem might also provide a solution to the
TBTF problem \cite{Ogiya}.

Some proposals have been suggested to solve the core-cusp problem. For
example, the existence of keV dark matter particles, as a candidate of
warm dark matter (WDM), has been proposed to solve the problem
\cite{Bode,Xue}.
However, recent observations indicate that the simplest model of
WDM (e.g. the non-resonant sterile neutrino model) cannot account for   
the major component of dark matter \cite{Abazajian,Viel,Seljak,Boyarsky}.
Some extra properties of WDM or free parameters are needed in order to    
satisfy the observational constraints. Some recent analyses even suggest
that WDM model cannot solve the core-cusp problem \cite{Maccio}. Another
proposal suggests
that core-like structures would be produced if dark matter particles are
self-interacting \cite{Spergel2}. Simulations show that
dark matter particles with a constant cross-section per unit mass
$\sigma/m \sim 1$ cm$^2$ g$^{-1}$ can produce core-like structures in
galaxies \cite{Burkert,Yoshida}.
Unfortunately, recent observations put a tight constraint on the
cross-section: $\sigma/m \le 1$ cm$^2$ g$^{-1}$ \cite{Randall,Zavala}. 
Therefore, it leaves only a small window open for this
velocity-independent self-interacting dark matter model to work
\cite{Zavala}. Nevertheless, this model might still endure if the
cross-section is velocity-dependent, though some more parameters have to 
be involved \cite{Loeb,Vogelsberger}.

Besides the above two proposals, some suggest that the energy exchange
between baryons and dark matter particles might also be possible to
produce core-like structures. These mechanisms include the
steller and supernova feedback \cite{Maccio2,Penarrubia}, and dynamical
friction \cite{ElZant,Tonini}. It
is now a controversial issue because these baryonic processes involve some
uncertainties, such as the total energy released by the supernovae
and the fraction of energy that can be transferred to the dark matter
haloes \cite{Ogiya}. For example, for a total mass of $10^9M_{\odot}$, recent studies show that at least $1/20$ of supernova energy is required to transfer to the dark matter halo to give $\gamma < 0.6$ \cite{Ogiya,Amorisco}. However, we are not sure whether this fraction of transferred energy is physically possible or not. Also, it is challenging to invoke baryonic  
processes as the main mechanisms to solve the core-cusp problem for some
dark-matter-dominated galaxies because the baryonic content is too small
to affect the dark matter distribution \cite{deBlok,Vogelsberger}.

In this article, I suggest another proposal that the core-cusp problem can
be solved by Sommerfeld-enhanced dark matter annihilation. The possibility
of the dark matter annihilation to solve the core-cusp problem is first 
suggested in \cite{Kaplinghat}. The required
annihilation cross-section is $<\sigma v> \sim
10^{-19}(m/\rm{GeV})$ cm$^{3}$ s$^{-1}$ \cite{Kaplinghat}. They suggest
two possible
mechanisms so that the required cross-section would not violate the relic
dark matter annihilation cross-section $<\sigma v>=3 \times 10^{-26}$  
cm$^3$
s$^{-1}$ \cite{Kaplinghat}. However, if the annihilation products are
visible particles such
as neutrinos, electrons or photons, this required cross-section is ruled  
out by observations \cite{Beacom,Madhavacheril}. Moreover, this model
predicts that halo core density is
universal ($\sim 0.02M_{\odot}$ pc$^{-3}$) \cite{Kaplinghat} while
observations indicate that the core density of dwarf
galaxies varies from $0.01-0.1~M_{\odot}$ pc$^{-3}$ \cite{Kormendy}. In
the following, I use the idea from \cite{Kaplinghat} but assume
that the dark matter annihilation is enhanced by the Sommerfeld's mechanism  
($<\sigma v> \propto v^{-\alpha}$) \cite{Sommerfeld,Yang}, and the   
annihilation products are invisible sterile neutrinos only. It can be
shown that a significant amount of dark matter particles would be
annihilated, which is enough to produce the observed core-like structures
in galaxies.

\section{The annihilation model}  
It has been suggested that dark matter would self-annihilate to give
smaller particles with high energy. The possible stable products formed
are photons, electron-positron pairs, and neutrinos. In particular, the 
fact that active neutrino have non-zero rest mass probably suggests that
right-handed neutrinos should exist, which may indeed be sterile
neutrinos \cite{Fan}. Some recent models suggest that dark matter    
particles can
annihilate dominantly into light dark neutrinos (sterile neutrinos) via
exchange of a Higgs field ($\chi \chi \rightarrow \Phi \Phi \rightarrow  
\nu_s \nu_s$) \cite{Cline}. This model can agree with the results
obtained from DAMA \cite{Bernabei} and CoGeNT \cite{Aalseth}
experiments, which point
toward light dark matter ($m \sim 1-10$ GeV) with isospin-violating and
possibly inelastic couplings \cite{Cline}. However, the light dark matter
model is largely constrained by observations, such as cosmic microwave  
background and SuperK limits. Cline and Frey (2012) \cite{Cline} propose a 
model of
quasi-Dirac dark matter, interacting via two gauge bosons, one of which
couples to baryon number and the other which kinetically mixes with the
photon. The annihilation product is dark neutrinos that do not mix
with the Standard Model. They also show that the dark neutrinos produced
in the universe would not violate the current observational bounds
\cite{Cline}.

In the following, we are going to discuss the consequences of the
sterile neutrinos being the only dark matter annihilation product (the 
model proposed in \cite{Cline} may be one of the possible scenarios). If 
the annihilation cross-section is large enough due to the Sommerfeld
enhancement, a significant amount
of dark matter would be changed to high energy sterile neutrinos. These
high energy sterile neutrinos would finally leave the structure and make
the central density lower. The Sommerfeld enhancement arises when a scattering object is coupled to a light mediator particle \cite{McDonald}. This enhancement can increase the cross-section for annihilation process in a velocity-dependent fashion due to the generic attractive force between the incident dark matter particles \cite{Yang}.

If dark matter particles were produced at the very
beginning, the annihilation cross-section should be close to $3 \times
10^{-26}$ cm$^3$ s$^{-1}$ \cite{Bertone}. However, the Sommerfeld
enhancement might significantly change the relic annihilation
cross-section to a lower value \cite{Zavala,Dent,Hannestad}. The actual
value of the annihilation cross-section at the thermal freeze out is
model-dependent. Nevertheless, \cite{Zavala} show that the Sommerfeld
enhancement near resonance would suppress the dark matter abundance by a
factor of a few. As a result, the annihilation cross section needs to be
suppressed by this same factor in order to be consistent with the observed 
relic density.

Therefore, we write the annihilation cross section at the thermal
freeze out as $<\sigma v_0>=1
\times 10^{-26}f$ cm$^3$ s$^{-1}$, where $f \sim 1$ is a model-dependent
parameter. Here, $v_0 \approx 0.2c$ is the
velocity of the dark matter particles at decoupling \cite{Armendariz}. 

If the velocity of dark matter particles $v$ is lower, the Sommerfeld 
enhancement might increase the cross-section to $<\sigma v>=<\sigma
v_0>(v_0^{\alpha}/v^{\alpha})$ \cite{Sommerfeld,Yang}. The value of
$\alpha$ is close to 1 for non-resonance case while $\alpha \approx 2$ for
resonance \cite{Zavala2}. Therefore, the low velocity
of dark matter particles in a galaxy or galaxy cluster would increase the
rate of annihilation to form core-like structures.

Also, this cross-section satisfies the unitarity bound. Harling and 
Petraki (2014) \cite{Harling} obtain the unitarity bound (upper bound) of 
the Sommerfeld-enhanced
cross-section, which is just a factor of 3 greater than the
unitarity bound without Sommerfeld enhancement. For $m \sim 1$ GeV and $v
\sim 10-1000$ km/s, the
unitarity bound is $<\sigma v>_{ub} \sim 10^{-13}-10^{-11}$ cm$^3$ 
s$^{-1}$ \cite{Harling,Griest,Mack}, which is much greater than the
cross-section considered in our model.

On the other hand, the presence of sterile neutrinos as a by-product
of dark matter annihilation would change the effective number of
neutrinos ($N_{eff}$) in cosmology. This number is strongly constrained by
cosmic microwave background anisotropies recently, which gives
$N_{eff}=2.88 \pm 0.20$ \cite{Rossi}. However, $N_{eff}$ depends
on the kinetic decoupling temperature of dark matter $T_{de}$, which is
a model-dependent parameter in cosmology. Assuming a reliable range
$T_{de} \sim 10-1000$ MeV (at $10^{-7}-10^{-3}$ s after Big Bang)
\cite{Cornell},
the number density of active neutrinos is $n_{\nu} \sim 10^{34}-10^{40}$
cm$^{-3}$. The number density of sterile neutrinos
produced from dark matter annihilation is given by $n_s \sim
\rho_{DM}^2<\sigma v_0>t_{de}/m^2 \sim 10^{23}-10^{32}$ cm$^{-3}$, where
$\rho_{DM}$ and $t_{de}$ are the mass density of dark matter and the age
of universe at the kinetic decoupling respectively. Therefore, the sterile
neutrinos produced at the kinetic decoupling are negligible compared with
the relic active neutrinos. Since the scale factor dependence are the same
for $n_s$ and $n_{\nu}$, the sterile neutrinos produced would not
significantly affect $N_{eff}$.

Since the mass of the mediator particle $m_{\phi}$ must be smaller than the dark matter mass \cite{McDonald,Feng}, this requires $m_{\phi} \sim 1$ MeV - 1 GeV. This new scalar field is likely to mix with the Higgs boson and possibly be in tension with current collider constraints. In view of this, a study in \cite{Cline2} examines this mixing effect seriously by assuming $m_{\phi} \sim 1$ GeV for the Sommerfeld enhancement. Since the mixing between the scalar field and the Higgs field involves some unknown free parameters, including $m_{\phi}$ and the mixing angle $\theta_{\phi h}$, a large area of parameter space is still possible to satisfy the constraints obtained in the Large Hadron Collider (LHC) experiments \cite{Cline2}. For the CMS measurements, there is an experimental upper limit for the Higgs total width of cross coupling. Based on the calculations in \cite{Cline2},  the upper limit of the Higgs-$\phi$ cross coupling $\lambda_1$ converges to about 0.05 for $m_{\phi}<1$ GeV, which is not a prohibited value.

Besides the Higgs constraints, the mediator particle would probably emit photons or charged bosons which would possibly conflict with the constraints from indirect detection experiments. However, the emission of photons by the mediator is model-dependent. For example, a study in \cite{Chen2} estimates a corresponding annihilation cross section by considering two dark matter particles annihilating via an intermediate pseudoscalar $A^0$ and a charged fermion $f$ in the loop (see the Feymann diagram in \cite{Chen2}). For a small coupling ($g \approx 1$), the model gives $<\sigma_{\gamma \gamma} v> \sim 10^{-40}$ cm$^3$ s$^{-1}$ $(m/1~{\rm GeV})^4 (500~{\rm GeV}/m_A)^4(500~{\rm GeV}/m_f)^2$, where $m_A$ and $m_f$ are the mass of the pseudoscalar particle and the fermion respectively \cite{Chen2}. For $m<1$ GeV, this cross-section is well-below any current constraint.

\section{Invisible dark matter annihilation in galaxies and galaxy
clusters}
In fact, numerical simulations should be performed in order to obtain a
precise picture of dark matter density profile $\rho(r,t)$ with
annihilation. However, as
we will see later, the annihilation is important only when the time $t$ is
sufficiently large ($t>1$ Gyr). Therefore, we can simply ignore the
dynamics of the
halo formation and assume that the initial dark matter density profile
follows the NFW profile \cite{Navarro}:
\begin{equation}
\rho(r,0)= \frac{\rho_sr_s^3}{r(r+r_s)^2},
\end{equation}
where $\rho_s$ and $r_s$ are scale density and scale radius of a
structure respectively. Numerical simulations show that $\rho_sr_s \approx
144(M/10^{12}M_{\odot})^{0.2}M_{\odot}$ pc$^{-2}$ \cite{Schaller}, where
$M$ is the total mass of a structure. The time evolution of the dark
matter density profile $\rho(r,t)$ is given by
\begin{equation}
\frac{\partial \rho(r,t)}{\partial t}=- \frac{[\rho(r,t)]^2<\sigma
v_0>v_0^{\alpha}}{m[v(r,t)]^{\alpha}}.
\end{equation}
Moreover, in a stable configuration, the circular velocity of dark
matter should depend on the mass profile $M(r,t)$ of a structure: $v(r,t)
\approx \sqrt{GM(r,t)/r}$. According to virial theorem and observations,
the velocity dispersion is $\approx v(r,t)$
\cite{Croton}. Therefore, we assume that the relative velocity of the
annihilating dark matter particles follows the velocity profile $v(r,t)$.

It is believed that most galaxies, including dwarf
galaxies, have a supermassive black hole with mass $M_{BH}$ at the center
\cite{McConnell,Reines,Seth}. Therefore we
write $M(r,t)=M_{BH}+M_{DM}$, where $M_{DM}=\int 4 \pi \rho(r,t)r^2dr$ is
the mass profile of dark matter. Here, we do not include the baryonic
component because we find that different functional forms of baryonic mass
profile just affect the resultant density profile slightly (for example, the difference between using an isothermal and a constant baryonic profile is less than 1\% in the resultant density profile). While in
galaxy clusters, we have
$M(r,t)=M_{DM}+M_b$, where $M_b$ is the mass profile of baryonic matter.
The baryonic component is important when $r$ is very small
because most of the dark matter is annihilated and there is no
supermassive black hole at the centre of galaxy cluster (except cD
clusters). In general, the locations of the galactic supermassive black
holes in a galaxy cluster are usually far away from the gravitational
centre of galaxy cluster (except cD clusters). Therefore, we set
$M_{BH}=0$ for small $r$ because they do not have any significant effect 
on the central density of galaxy cluster. The
baryonic component can be simply modelled by a constant density profile   
for small $r$ \cite{Chen}. It is significant only when $r \le 0.1$ kpc.

Since the velocity profile is also time-dependent, Eq.~(2) has to be
solved numerically. Nevertheless, we can discuss some important features
and behaviours of the resulting dark matter density profile by using some
analytic approximations, especially for small $r$ regime.

In a typical galaxy, we have $M_{BH} \gg M_{DM}$ at small $r$. Therefore,
the total mass profile is time and radius independent ($M(r,t)
\approx M_{BH}$) near the galactic centre. In this regime, Eq.~(2) gives 
\begin{equation}
\rho(r,t)=\left[ \frac{1}{\rho(r,0)}+ \frac{<\sigma
v_0>v_0^{\alpha}r^{\alpha/2}t}{(GM_{BH})^{\alpha/2}m}
\right]^{-1}=\left( \frac{r}{\rho_sr_s}+Kr^{\alpha/2}t
\right)^{-1}.
\end{equation}
If $t$ is sufficiently large, $\rho(r,t)$ would follow $r^{-\alpha/2}$ at
small $r$. When $M_{DM}$ dominates the mass profile, the total mass
would be time and radius dependent. Nevertheless, if we neglect the
time-evolution factor (it is a good approximation only near the core
radius, and when $t$ is small), we can get an approximate analytic
expression by using $M(r,t) \approx 2 \pi \rho_sr_sr^2$ for $r<r_s$.
Eq.~(2) gives  
\begin{equation}
\rho(r,t)=\left[ \frac{1}{\rho(r,0)}+ \frac{<\sigma  
v_0>v_0^{\alpha}t}{(2 \pi G \rho_sr_sr)^{\alpha/2}m}  
\right]^{-1}=\left( \frac{r}{\rho_sr_s}+K'r^{-\alpha/2}t
\right)^{-1}. 
\end{equation}

Moreover, since $M_{BH}=0$ for small $r$ in most of the galaxy
clusters, the
baryonic component will dominate the mass profile for very small $r$ when 
most of the dark matter near the centre is annihilated ($t$ is 
sufficiently large). In this case, the total mass profile is just 
radius-dependent. We have
\begin{equation}
\rho(r,t)=\left[ \frac{1}{\rho(r,0)}+ \frac{<\sigma  
v_0>v_0^{\alpha}r^{\alpha/2}t}{(1.33 \pi G \rho_br^3)^{\alpha/2}m}
\right]^{-1}=\left( \frac{r}{\rho_sr_s}+\tilde{K}r^{-\alpha}t
\right)^{-1}, 
\end{equation}
where $\rho_b$ is the central baryon density. Since $\alpha>0$, from
Eqs.~(4) and (5), we notice that $\rho(r,t)$ will rise again towards small
$r$ for both galaxies and galaxy clusters.

From Eq.~(4), we can see that there exists a turning point in
the density when $K'r^{-\alpha/2}t \approx r/\rho_sr_s$ for dwarf and
normal galaxies. Therefore, the core radius of
the structure $r_c$ should be the same order of magnitude as the position
of the turning
point. For $\alpha=1$ and taking $\rho_sr_s=(58-144)M_{\odot}$
pc$^{-2}$ for dwarf and normal galaxy scale \cite{Schaller}, we
have $r_c \sim (10^{-4}-10^{-3})(m/\rm 1~GeV)^{-2/3}$ kpc, which is too
small to be the
core radius. For $\alpha=2$, we get $r_c \sim (0.1-1)(m/\rm 1~GeV)^{-1/2}$
kpc, which generally agrees with observations. This shows that
our model is consistent only with $\alpha=2$.

Let's assume that $m=1$ GeV, $f=1$ and $\alpha=2$. By solving Eq.~(2)
numerically with small time-steps, we can generate the time
evolution of $\rho(r,t)$ for a typical dwarf galaxy, normal galaxy and   
galaxy cluster, respectively (see Figs.~1-3). All the parameters used are
listed in Table
1. We use the empirical formula from \cite{McConnell} to model the
corresponding mass of supermassive black hole in a typical dwarf galaxy
and in a normal galaxy. We can notice from the figures that the density
fluctuates at small $r$. This is consistent with the features
obtained from the approximate analytic expressions in Eqs.~(3)-(5).
The inner slope is much shallower than the NFW profile. Secondly,
neglecting the fluctuations of density for small $r$, we approximate the
resultant density profile with a cored-NFW profile
$\rho=\rho_c(1+r/r_c)^{-1}(1+r/r_s)^{-2}$ in each case. The core radii
are $r_c \sim
1$ kpc and $r_c \sim 10$ kpc for galaxy and galaxy cluster respectively.
The small core radii in dwarf galaxy and normal galaxy
generally agree with observations \cite{Donato,Kormendy}. However, the
structure of core in galaxy cluster is not obvious and the core
radius is too small for us to observe. Current
observational data from galaxy clusters for $r<100$ kpc is highly
uncertain. This may explain why core-like structures are not commonly
found in galaxy clusters, even though they might exist.

\begin{table}
\caption{The parameters used in our model.}
 \label{table1}
 \begin{tabular}{@{}lcccc}
  \hline
  Structure & $\rho_s$(g cm$^{-3}$) & $r_s$(kpc) & $M_{BH}(M_{\odot})$ &
$M(M_{\odot})$ \\
  \hline
  Dwarf galaxy & $10^{-24}$ & $3.2$ & $10^4$ & $10^{10}$ \\
  Normal galaxy & $3 \times 10^{-25}$ & $32$ & $10^6$ & $10^{12}$ \\   
  Galaxy cluster & $7 \times 10^{-26}$ & $500$ & $0$ & $10^{15}$ \\
  \hline
 \end{tabular}
\end{table}

The resulting density profile in our model is somewhat different from
the result obtained from \cite{Kaplinghat}. It is because the
cross-section used in
\cite{Kaplinghat} is constant while the cross-section in this model is
velocity-dependent. Fig.~4 shows a plot of the velocity profiles of
dark matter particles in a typical galaxy and a typical dwarf galaxy at
$t=12$ Gyr. We can
notice that the drop and rise in the resulting density profile strongly
correlates with the velocity profile of dark matter. If there is no
supermassive black hole or baryon in the structure (i.e. $M_{BH}=0$ and
$M_B=0$), the resulting density profile can be approximated by Eq.~(4) 
only
(density decreases towards the centre), and the shape of the density
profile would be similar to the one shown in Fig.~3 because the effect of
baryon in galaxy cluster is nearly negligible. If there is no
Sommerfeld enhancement (i.e. $\alpha=0$), from Eq.~(4), the resulting
central density profile would be constant ($\rho(t) \approx m/<\sigma
v_0>t$), which reduces to the result obtained from \cite{Kaplinghat}.

On the other hand, Fig.~5 shows that core-like structure disappears when
$M_{BH}$ is large ($M_{BH} \sim 10^8M_{\odot}$). Since large $M_{BH}$   
usually correspond to large galactic mass \cite{Bandara}, this suggests
that
core-like structures can be easily found in dwarf galaxies but not in  
large galaxies (with mass greater than $10^{13}M_{\odot}$).

If we take the local maximum turning point of the density profile being
the core radius $r_c$, the mass profile at $r=r_c$ would be dominated by
dark matter. At the turning point, from Eq.~(4), we have $r_c \sim  
\sqrt{\rho_sr_sK't_0}$, where $t_0 \approx 12$ Gyr is the age of a
galaxy.
The central density is given by $\rho_c \approx \rho(r_c,t_0) \approx
\rho_sr_s/r_c$. Therefore, we have $\rho_cr_c \approx \rho_sr_s$. As 
$\rho_sr_s \propto M^{0.2}$, our result is consistent with the
empirical fits from observational data $\rho_cr_c \propto M^{0.2}$
\cite{DelPopolo}. Since this is a
slow varying function of $M$, the product $\rho_cr_c \sim 100M_{\odot}$ 
pc$^{-2}$ is nearly a constant
for dwarf galaxies and normal galaxies, which agrees with the
empirical fits using some cored-density profiles
$\rho_cr_c=141^{+82}_{-52}M_{\odot}$ pc$^{-2}$ \cite{Donato,Gentile}. For
$M \sim 10^{10}M_{\odot}$ in dwarf galaxies, our result gives $\rho_c
\approx 0.07M_{\odot}$ pc$^{-3}$, which matches the average central
density ($\sim 0.01-0.1M_{\odot}$ pc$^{-3}$) in dwarf galaxies
\cite{Kormendy}.

Moreover, since many dark matter particles are annihilated, the central
density become much smaller. This would provide a possible solution to
solve the TBTF problem \cite{Ogiya}. For $r<250$ pc, the dark matter
density is less than $10^{-23}$ g cm$^{-3}$ in a dwarf galaxy. Therefore,
the rotational velocity would be less than 13 km/s, which can
satisfactorily address the TBTF problem \cite{Walker,Wolf}.

\begin{figure}
 \includegraphics[width=82mm]{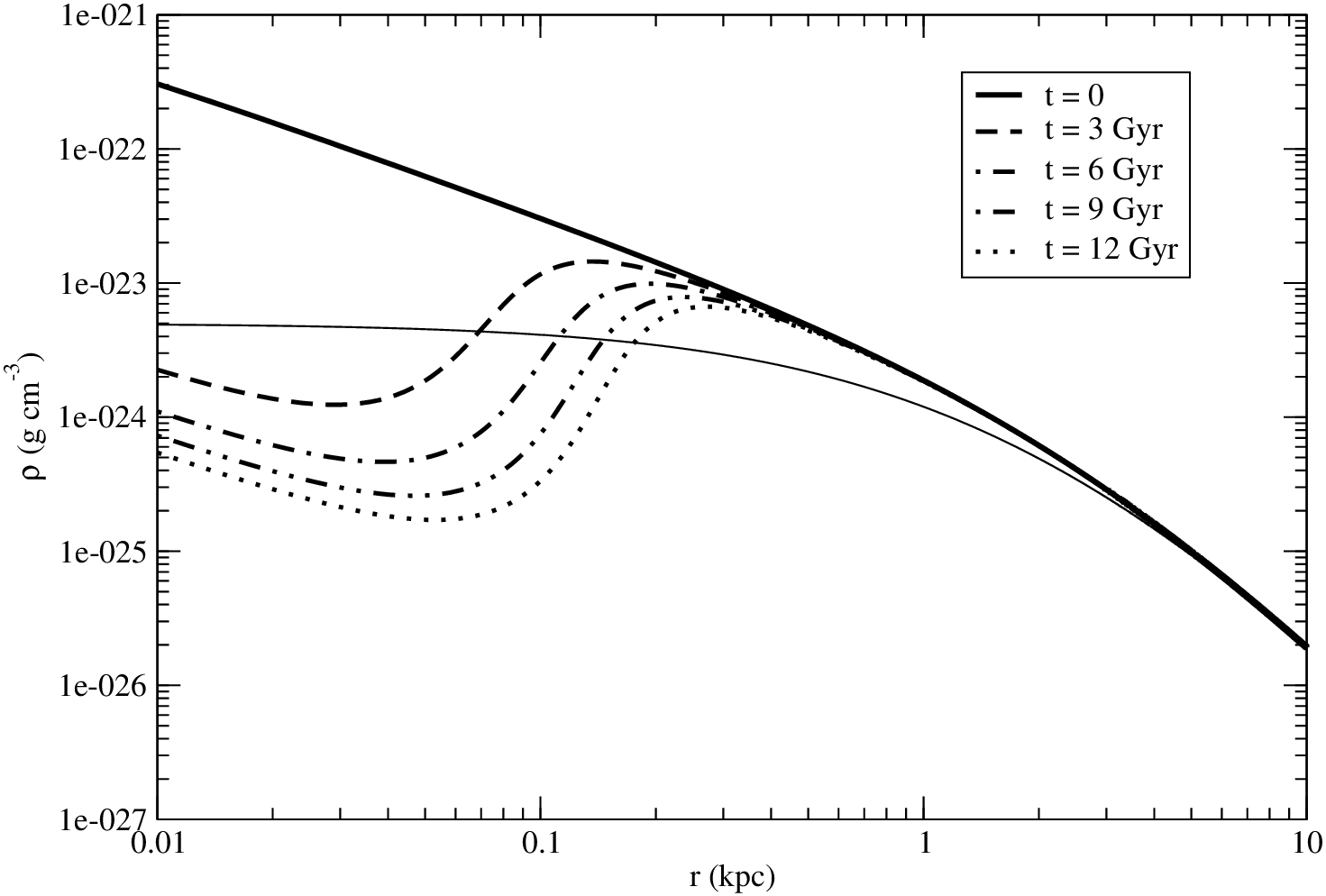}
 \caption{Time evolution of dark matter density profile in a dwarf galaxy.
The thin solid line represents a cored-NFW density profile with
$\rho_c=5 \times 10^{-24}$ g cm$^{-3}$ and $r_c=0.7$ kpc.}
\vskip 10mm
\end{figure}

\begin{figure}
 \includegraphics[width=82mm]{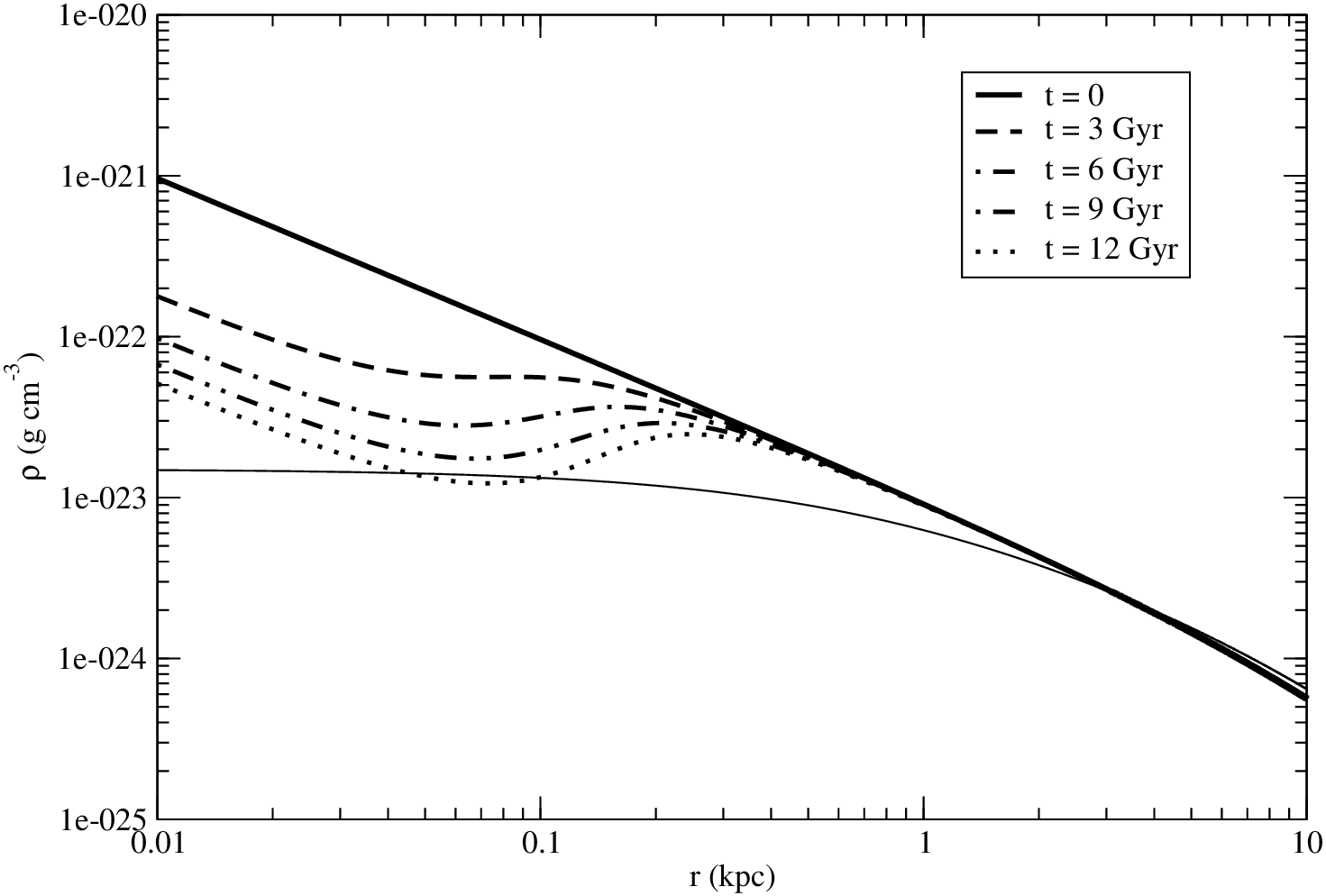}
 \caption{Time evolution of dark matter density profile in a normal   
galaxy. The thin solid line represents a cored-NFW density profile
with $\rho_c=1.5 \times 10^{-23}$ g cm$^{-3}$ and $r_c=0.8$ kpc.}
\vskip 10mm
\end{figure}

\begin{figure}
 \includegraphics[width=82mm]{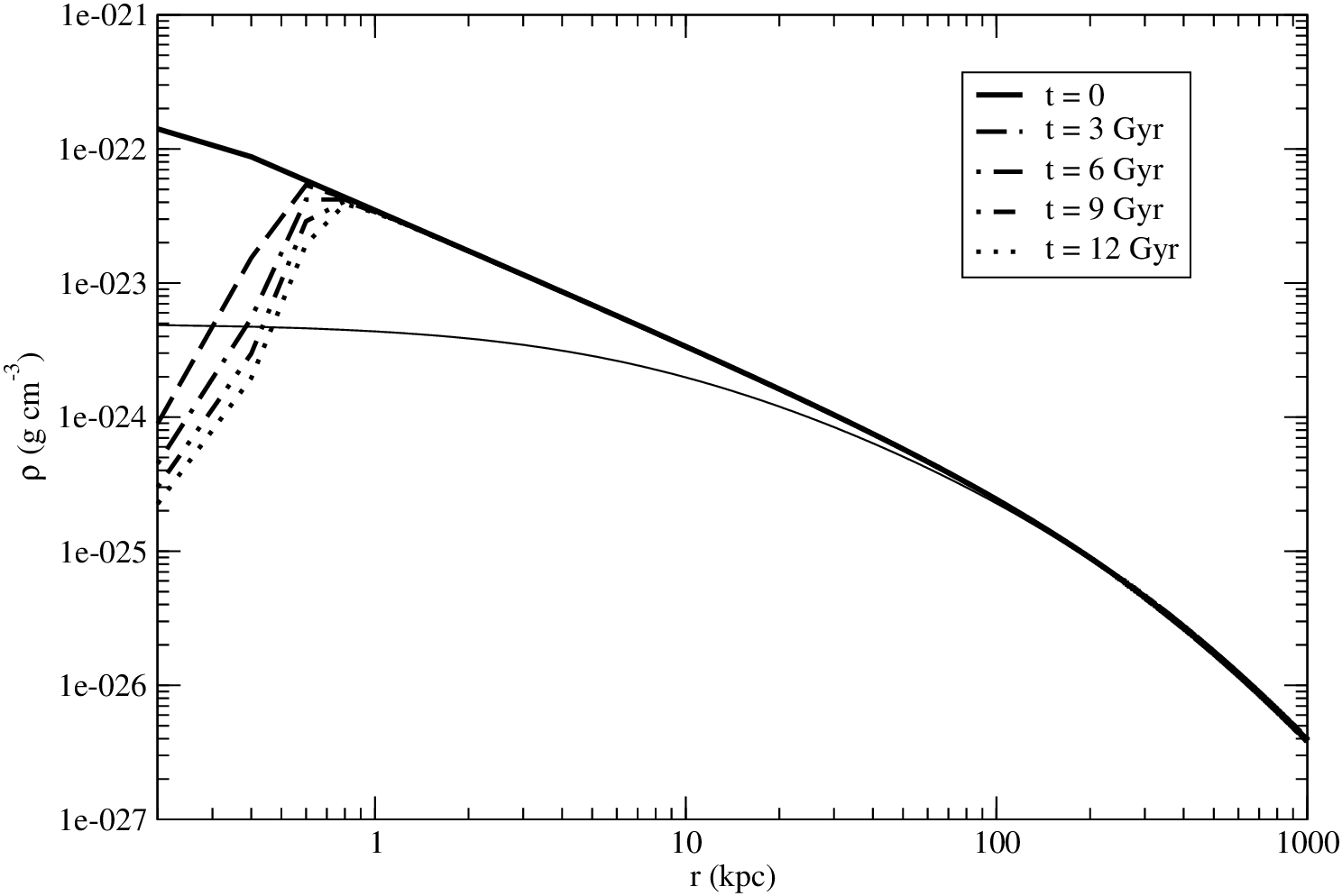}
 \caption{Time evolution of dark matter density profile in a galaxy
cluster. The thin solid line represents a cored-NFW profile with
$\rho_c=5 \times 10^{-24}$ g cm$^{-3}$ and $r_c=7$ kpc.}  
\vskip 10mm
\end{figure}

\begin{figure}
 \includegraphics[width=82mm]{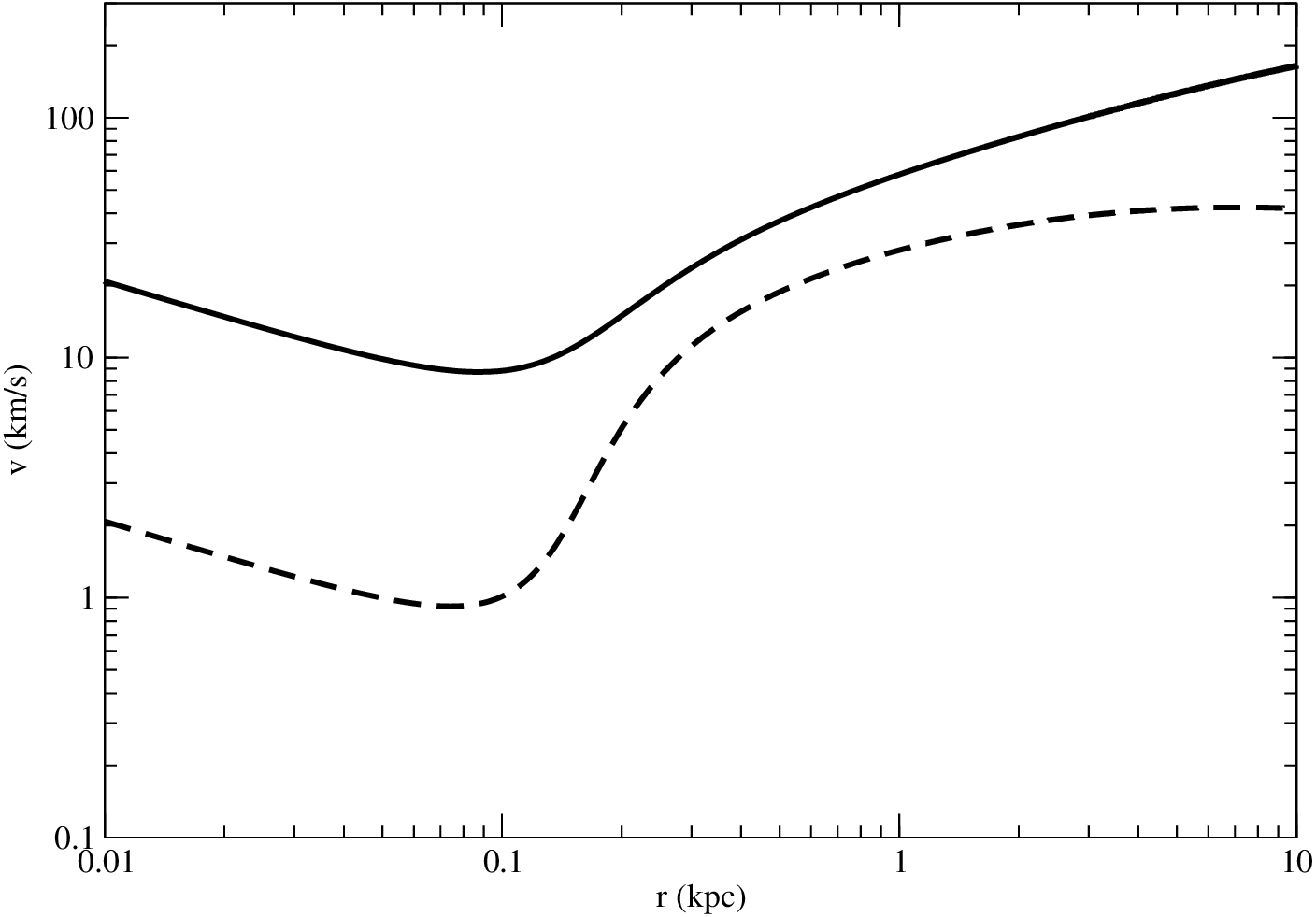}  
 \caption{The circular velocity profile of dark matter in a typical galaxy
(solid line) and a typical dwarf galaxy (dashed line) at $t=12$ Gyr.}
\vskip 10mm
\end{figure}

\begin{figure}
 \includegraphics[width=82mm]{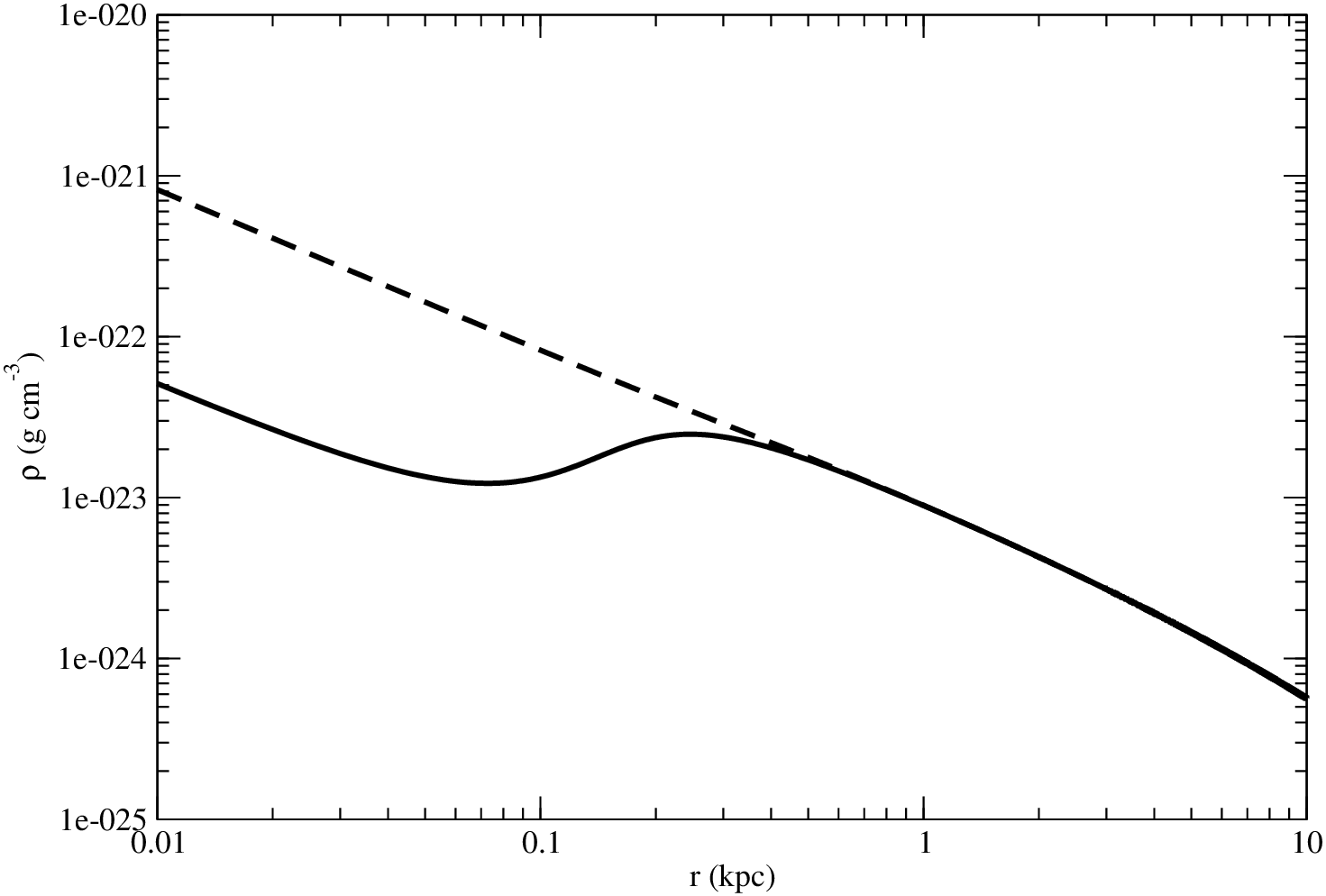}
 \caption{The dark matter density profile $\rho(r,t_0)$ in a normal galaxy
with different supermassive black hole mass (solid line:
$M_{BH}=10^6M_{\odot}$; dashed line: $M_{BH}=10^8M_{\odot}$).}  
\vskip 10mm
\end{figure}

\section{Discussion}
In this article, I show that the Sommerfeld-enhanced invisible dark matter
annihilation is able to solve the core-cusp problem and obtain some
general features that agree with observations. In order to produce the
observed core size in galaxies, the value of $\alpha$ should be around 2,
which means that the Sommerfeld boost is very close to resonance. Since
the annihilation products are invisible sterile neutrinos, the
Sommerfeld-enhanced
cross-section basically does not violate any observational bounds. This
model can also give a satisfactory explanation why core-like
structures commonly found in dwarf galaxies, but not in galaxy clusters.  
It is because the core size is too small to be observed 
in galaxy clusters. Moreover, this model just involves
only two free parameters, the mass of supermassive black hole $M_{BH}$,
and the mass of a dark matter particle $m$, which
is assumed to be 1 GeV (the free parameter $f$ can combine with $m$ to
become a single parameter $f/m$). In general, a smaller value of $m$ would
give a more obvious core-like structure and a lower central density. Since
our result is consistent with the typical value of central density in
galaxies, the value of $m$ should be of the order 1 GeV.

Since a significant amount of dark matter would be annihilated,
the dark matter content would be smaller in the inner region. Based on our
calculations, more than 60\% of the dark matter would be annihilated
within 1 kpc. Surprisingly, this
agrees with recent observations that the dark matter content is close to
zero for the central part of galaxies \cite{Lelli}. Some observations    
also
reveal that the dark matter content decreases with decreasing radius
\cite{Kassin}, which could also be explained by a higher rate of dark 
matter annihilation in the inner radius.

On the other hand, it is possible to have a very tiny baryonic branching  
ratio of the annihilation such that some photons or electron-positron
pairs are directly created. However, the branching ratio should be less than  
$10^{-8}$ for these visible channels in order to satisfy the
observational bounds \cite{Madhavacheril}.

In our model, although the inner slope of the density is much shallower   
than the NFW profile, the inner central density is not really a constant.
However, due to observational uncertainties, the small fluctuations in the
inner small region are usually smoothed out by using a constant density 
profile. If we have some good techniques in the future that can precisely 
probe the central
density of dark matter, our model can be severely tested by three ways.
First, the inner dark matter density is smaller than the NFW profile and
goes like $1/r$ when $M_{BH}$ dominates the central mass. Second, there
is a peak and trough in the dark matter density
profile near the core radius. Third, our model predicts that core-like    
structures disappear in galaxies with large supermassive black holes.
These features would be able to detect if we can precisely obtain the 
baryon density profile in galaxies or galaxy clusters in the future.

\section{acknowledgements}
This work is partially supported by a grant from the Hong Kong Institute of Education (Project No.:RG57/2015-2016R).





\end{document}